\documentclass[conference]{IEEEtran}
\usepackage[utf8]{inputenc}
\usepackage{cite}
\usepackage[margin=1.1in]{geometry}
\usepackage{graphicx}
\usepackage{lipsum}
\usepackage{amsmath}
\usepackage{ amssymb }
\usepackage{ wasysym }
\usepackage{dsfont}
\usepackage{xcolor}
\usepackage{authblk}

\usepackage[makeroom]{cancel} 
\usepackage{dirtytalk}
\usepackage{charter}    
\usepackage{subcaption}  
\usepackage{hyperref}
\usepackage{physics}
\usepackage[labelsep=period]{caption} 
\usepackage[normalem]{ulem} 

\usepackage{tabularx}

\newcounter{scheme}
\newenvironment{scheme}
  {\par\addvspace{\topsep}
   \noindent
   \tabularx{\linewidth}{@{} X @{}}
    \hline
    \refstepcounter{scheme}\textbf{Network Generation Routine} \\
    \hline}
  { \\
    \hline
   \endtabularx
   \par\addvspace{\topsep}}

\newlength{\myspace}
\DeclareUnicodeCharacter{0301}{\'{}}

\begin{document}

\title{Quantum Key Distribution over \\Complex Networks}

\author[1,2]{Luca Mariani}
\author[3]{Raja Yehia}
\author[3]{Carlos Pascual-García}
\author[3,4]{Federico Centrone}
\author[5,6,7]{\\Jasper van der Kolk}
\author[5,6,8]{M. \'Angeles Serrano}
\author[3,8]{Antonio Acín}

\affil[1]{ICAR CNR - Institute for High Performance Computing and Networking, via Pietro Bucci, Rende, Italy}
\affil[2]{University of Salerno, Department of Physics “E. R. Caianiello”, via Giovanni Paolo II 132, Fisciano, Italy}
\affil[3]{ICFO - Institut de Ciencies Fotoniques, The Barcelona Institute of Science and Technology, Castelldefels, Spain}
\affil[4]{Universidad de Buenos Aires, Instituto de Física de Buenos Aires (IFIBA), Ciudad Universitaria, 1428 Buenos Aires, Argentina.}
\affil[5]{University of Barcelona Institute of Complex Systems (UBICS), E-08028 Barcelona, Spain}
\affil[6]{Department of Condensed Matter Physics, University of Barcelona, Martí i Franquès 1, E-08028 Barcelona, Spain}
\affil[7]{Department of Network and Data Science, Central European University Vienna, Vienna 1100, Austria}
\affil[8]{ICREA - Institució Catalana de Recerca i Estudis Avançats, 08010 Barcelona, Spain}

\maketitle
\begin{abstract}
    There exist several initiatives worldwide to deploy quantum key distribution (QKD) over existing fibre networks and achieve quantum-safe security at large scales. To understand the overall QKD network performance, it is required to transition from the analysis of individual links, as done so far, to the characterization of the network as a whole. In this work, we undertake this study by embedding QKD protocols on complex networks, which correctly model the existing fiber networks. We focus on networks with trusted nodes and on continuous-variable (CV) schemes, which have much higher key rates than their discrete-variable (DV) counterparts. In the effective CV network, however, many of the unique properties of complex networks, such as small-worldness and the presence of hubs, are lost due to the fast decay of the key rate with physical distance for CV systems. These properties can be restored when considering a hybrid network consisting of both CV and DV protocols, achieving at the same time high average rate and inter-connectivity. Our work opens the path to the study of QKD complex networks in existing infrastructures. 
\end{abstract}

The security of existing encryption protocols such as RSA~\cite{RSA_1978} is compromised by quantum computers, as quantum algorithms can break such schemes efficiently~\cite{ShorAlg}. To address this threat and design protocols secure against quantum computers, two alternatives exist: post-quantum cryptography, where protocols base their security on computational problems believed to be hard even for quantum computers~\cite{bernstein2017post}; or quantum cryptography, whose security follows from the laws of quantum physics~\cite{pirandola2020advances}. Quantum key distribution (QKD)~\cite{BB84, E91, BBM_1992} is the most advanced quantum cryptography application that enables two distant, honest parties denominated Alice and Bob to generate a shared secret key. The security of this key against any potential eavesdropper, typically called Eve, is based on the principles of quantum mechanics. 

To attain quantum-safe security at large scales, several initiatives worldwide, such as the European Quantum Communication Infrastructure (EuroQCI~\cite{euroQCI}), have been launched in recent years to deploy QKD over existing fiber networks. It is therefore timely and necessary to analyze the \emph{collective} properties of QKD networks with a large number of users to understand and guide efforts on QKD deployment. In this work, we address this question within the framework of complex network theory~\cite{NetworkScienceAlbertLaszlo,Serrano_Bogunna_2022,Newman2010Networks}, a well-established branch of network science that enables the study of principles underlying the structure and behavior of networks with non-trivial connectivity features. Existing fibre networks, over which QKD is being deployed, are examples of complex networks. While there exist a few previous works connecting complex networks and quantum information protocols~\cite{Perseguers2010,britoStatisticalPropertiesQuantum2020, britoSatelliteBasedPhotonicQuantum2021, centroneCostRoutingContinuousvariable2023, mengPathPercolationQuantum2024,Cirigliano2024Optimal}, studies of QKD performance on realistic complex networks~\cite{nokkala2023complex} are still missing. We bridge this gap by embedding QKD protocols in complex network models of the classical Internet~\cite{faloutsosPowerlawRelationshipsInternet1999,MASerrano2005Completition,MASerrano2006Modeling}. We aim at deriving rules to design networks that optimize the overall QKD performance, and understand the impact of intrinsic complex network properties. On the other hand, from a fundamental perspective, it is interesting to identify and characterize critical phenomena in the resulting QKD network. \\

\begin{figure*}[t!]
    \centering
    \begin{subfigure}[b]{0.49\textwidth}
        \centering
        \includegraphics[width=1\textwidth]{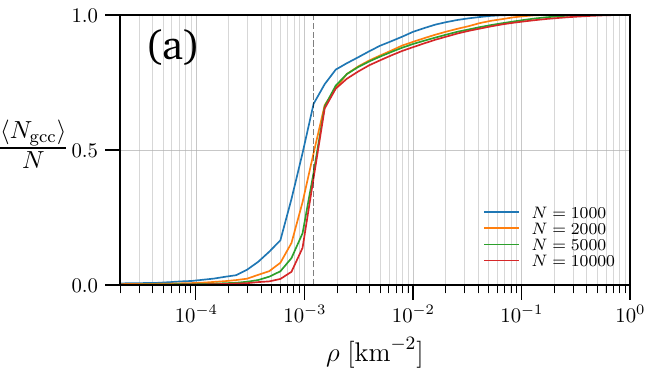}
        \phantomcaption
        \label{fig:connectivities_vs_rho}
    \end{subfigure}
    \hfill
    \begin{subfigure}[b]{0.49\textwidth}
        \centering
        \includegraphics[width=\textwidth]{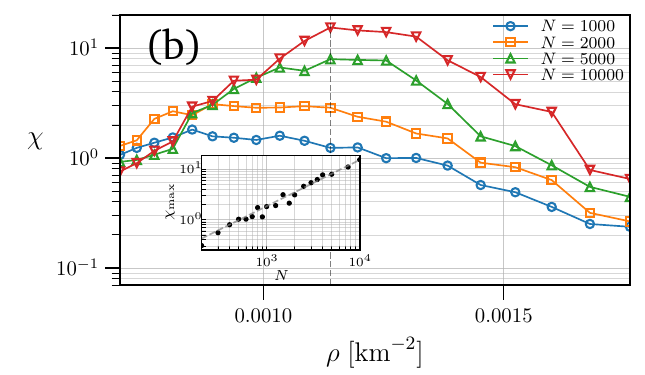}
        \phantomcaption
        \label{fig:susceptibility_curves}
    \end{subfigure}
    \caption{(a) 
    Network connectivity as a function of the density $\rho$ of nodes in real space, for several numbers $N$ of nodes in the system. The dashed vertical line corresponds to the critical density $\rho_c$, which is the percolation threshold. Connectivity is defined as the ratio of the average size $N_\mathrm{gcc}$ of the largest connected component over the size $N$ of the whole network (see App.~\ref{app:networkdef} for the precise definitions of complex network properties). (b) The main plot shows the susceptibility $\chi\equiv \frac{\langle (N_\mathrm{gcc} - \langle N_\mathrm{gcc} \rangle)^2 \rangle}{\langle N_\mathrm{gcc} \rangle}$ as a function of $\rho$ for different sizes of the network. The inset shows the height $\chi_\mathrm{max}$ of the peak, computed for a range of $N$. For this specific plot, and to better resolve the positions of the peaks, we collected data points ranging within a narrower interval of densities, centered around the percolation threshold, and averaging the curve over 40 instances of the network.}
    \label{fig:phase_transition}
\end{figure*}

The mathematical tool used to represent networks and analyze their properties is the \textit{graph}, namely a set of \textit{nodes} (or \textit{vertices}) and \textit{links} (or \textit{edges}) that connect them. Nodes are an abstraction of the agents in the system and edges represent the relationships between them. In our case, nodes are honest users willing to establish secret keys and edges represent the connecting fibers. 

To describe the network over which QKD will be deployed we use techniques from \emph{Network Geometry}, see~\cite{Serrano_Bogunna_2022,bogunaNetworkGeometry2021} and App.~\ref{app:network}. This framework allows one to study complex networks and, most importantly, explain the emergence of their paradigmatic properties in real systems. There, nodes are assumed to live in a latent geometric space that conditions the network topology. Nodes that lie close to each other in this space are said to be more \emph{similar} and therefore more likely to be connected. The underlying metric space is therefore often called the \emph{similarity space}. In addition to similarity, another important concept in the formation of complex networks is \emph{popularity}: more popular nodes will be more likely to form connections~\cite{barabasiEmergenceScalingRandom1999}. In this article, we employ the $\mathbb{S}^D$-model~\cite{SelfSimilaritySerrano2008,SmallWorldsBoguna2020}, which takes into account both similarity and popularity dimensions. The details of the model can be found in App.~\ref{sec:geomodel}, but it is important for our purposes that this model has an explicit geometric component, represented by a $D$-dimensional sphere as the similarity space, where we assume nodes to be uniformly distributed.

In this work, we develop a routine to numerically simulate the behavior of a quantum secure network, starting from a graph generated with the $\mathbb{S}^D$-model. Once the graph is constructed, the coordinates of the nodes in the geometric latent space are also interpreted as coordinates in the physical space. In this way, after obtaining the distances between the nodes, we can associate a key rate $K$ with each edge, see also App.~\ref{App:qkd}. The dependence of the key rate on distance implies the existence of a threshold over which no positive secret key rate is achievable. Links with distances larger than this critical value are useless and can be removed from the QKD network, in a process that we refer to as \textit{pruning}. We can then analyze the properties and QKD performance of the resulting complex network for a varying density of nodes $\rho$ in the physical space. The details of this routine can be found in App.~\ref{App:NetworkGen}, the definition of the different figures of merit are given in App.~\ref{App:figures_merit} and the parameters used in the simulation are explained in App.~\ref{app:AnalysisQNet}.

To simplify the analysis, we focus on the asymptotic regime and quantify the QKD rate $K$ using the Devetak-Winter bound~\cite{devetak2005distillation}. The exact expression for $K$, which is detailed in App.~\ref{App:qkd}, depends on the physical distance, the standard parameters in fiber communications, and the protocol considered. The dependence on distance comes mainly from exponentially growing losses in optical fibers. In our analysis, we first consider a continuous-variable (CV) protocol, which yields high rates at metropolitan scales, can be implemented using standard telecom technologies, and is easier to integrate in existing infrastructures. However, the performance of CV approaches declines rapidly as the average distance between users increases. We thus incorporate into the model the option of using a discrete-variable (DV) protocol which, at the cost of a reduced key rate, tolerates higher losses.
\\

Let us now examine the properties of the QKD networks generated by our routine. As illustrated by Fig.~\ref{fig:connectivities_vs_rho}, as $\rho$ grows, the topology of CV networks after the pruning gradually transforms from a fully disconnected graph to the original network, consisting of one dominant giant component. This is a minimum requirement for secure communication between any two nodes. This change of topology happens in a quite narrow interval of $ \rho $, suggesting the presence of an explosive percolation transition~\cite{da_costa_explosive_2010,dsouza_anomalous_2015}. This hypothesis is corroborated by the study of the susceptibility $\chi$, which quantifies the amplitude of the fluctuations in the size of the giant component~\cite{ferreiraEpidemicThresholdsSusceptibleinfectedsusceptible2012, castellanoNumericalStudyPercolation2016}. As shown in Fig.~\ref{fig:susceptibility_curves}, the susceptibility exhibits a peak that becomes sharper as $N\rightarrow \infty$, a key indicator of a continuous phase transition. The percolation threshold $\rho_c$ can thus be estimated by looking at the density at which $\chi$ reaches its peak, approximately equal to $1.1\times 10^{-3}\ \mathrm{km}^{-2}$.
This value gives an estimate of the maximum spacing allowed between nodes of the original network, which results in a CV-QKD network with a giant connected component after pruning. In this configuration, each node has its nearest neighbor at an average distance of roughly 30 km. \\

\begin{figure}[h!]
    \centering
    \begin{subfigure}[b]{0.49\textwidth}
        \centering
    \includegraphics[width=0.99\textwidth]{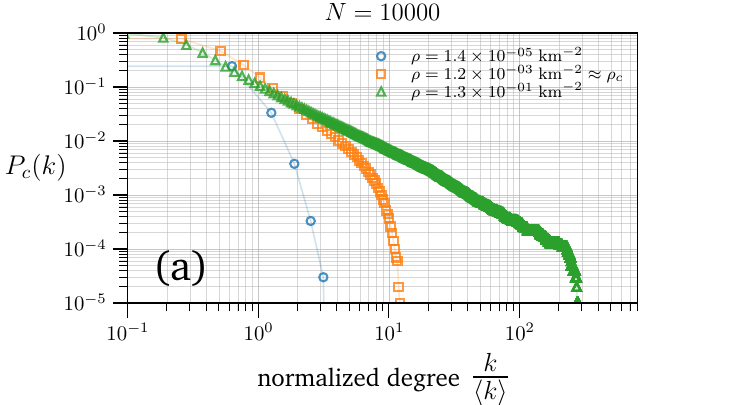}
        \phantomcaption
        \label{fig:deg_dist}
    \end{subfigure}
     \vspace{-15pt}
\begin{subfigure}[b]{0.49\textwidth}
        \centering
        \includegraphics[width=0.99\textwidth]{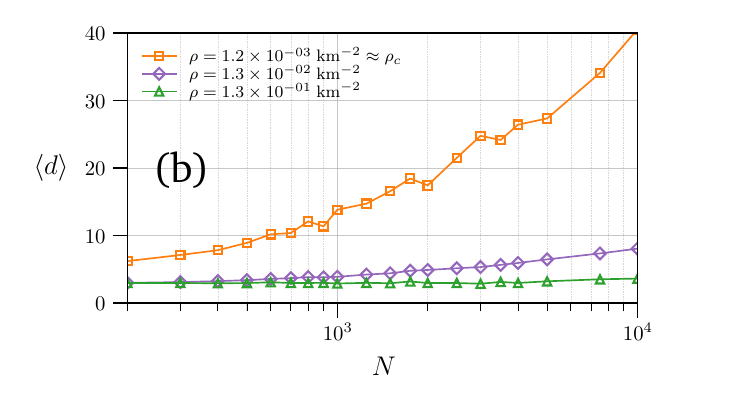}
        \phantomcaption
        \label{fig:distance_vs_logN}
    \end{subfigure}
    \vspace{-15pt}
 \begin{subfigure}[b]{0.49\textwidth}
        \centering
        \includegraphics[width=0.99\linewidth]{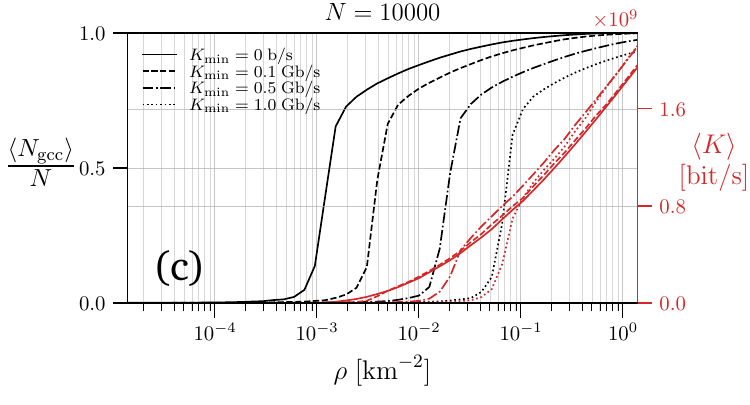}
        \phantomcaption
        \label{fig:threshold_rate}
    \end{subfigure}
    \caption{(a) Complementary cumulative degree distribution $P_c(k)$ of the pruned network for three different densities, plotted against the degree normalized by its average $\frac{k}{\langle k \rangle}$. The three values for $\rho$ are chosen to be significantly above (green triangles), below (blue circles) or approximately equal (orange squares) to the percolation threshold. This plot is done with a number of nodes $N=10000$, averaged over 10 instances of the network. The term $P_c(k)$ represents the fraction of nodes in the network with more than $k$ neighbors. Normalizing the degree by its average allows for a fair comparison between curves with different $\langle k \rangle$ and highlights the loss of the power-law behavior at low node densities. (b) Average topological distance $\langle d \rangle$, as a function of the size $N$ of the network. The topological distance $d$ between two nodes is the number of edges in the shortest path, if any, that connects them. For each point, the average is taken over all possible node pairs in 10 realizations of the network, and is well-defined only for percolated systems, where most nodes belong to the same component. In both plots, different curves correspond to different orders of magnitude of the node density. (c) Connectivity (black lines) and average rate (red lines) of the network for different values of the rate threshold $K_\mathrm{min}$. }
    \label{fig:topology}
\end{figure}

A feature that is commonly found in real complex networks, and reproduced by the $\mathbb{S}^D$ model, is a power-law degree distribution. It implies the presence of \textit{hubs}, i.e. nodes having a very large number of neighbors. We see in Fig.~\ref{fig:deg_dist} that this feature is fundamentally preserved after pruning in densely populated systems ($\rho \gg \rho_c$, e.g. green curve), where the density of points is so large that almost no edges are removed. For sparser systems, however, the tail of the degree distribution is cut, meaning that the number of hubs drops dramatically. This is a direct consequence of imposing a maximal distance between two nodes due to CV-QKD constraints, which reduces substantially the amount of available neighbors. Such a cutoff not only prevents the formation of hubs but also leads to longer path lengths, as evident from the analysis of the average topological distance $\langle d \rangle$ (Fig.~\ref{fig:distance_vs_logN}). Densely packed networks are, again, barely affected by pruning. They exhibit the small-world property, that famously characterizes the classical Internet: a path with a very low number of links is sufficient to connect any two nodes in the network (green curve). Instead, as $\rho$ decreases, long-range connections are gradually excluded from the system. Without those shortcuts, routes connecting two distant nodes in a CV-QKD network architecture are segmented into multiple short-range links, causing $\langle d \rangle$ to grow faster with $N$ (orange curve).

Finally, we study the average key rate $\langle K \rangle$ to analyze the performance of the system from a cryptographic standpoint. 
For any two nodes $A$ and $B$ in the network, the achievable key rate between them, $K_{AB}$, is determined by a pathfinding algorithm. This algorithm selects, among all available paths connecting $A$ and $B$, the one that minimizes the time required for key generation over all the intermediate links forming the path (see App.~\ref{App:figures_merit} for further details). Then, $\langle K \rangle$ is calculated as the average over several choices of $A$ and $B$ and over different network realizations.

As expected, $\langle K \rangle$ strongly depends on the connectivity of the network: communication is only possible for networks that reach a density $\rho>\rho_c$ large enough to guarantee percolation and the emergence of a single connected component in the pruned network. The behavior of the average rate is influenced by the interplay of two dependencies: First, when $\rho$ is high, nodes are closer to each other, so the one-to-one rates across single edges are on average larger. Second, these networks provide better interconnection, thereby offering a broader range of options for the path between any two nodes, which also strengthens the resilience of the network. 

In Fig.~\ref{fig:threshold_rate}, we show $\langle K \rangle$ for different network densities, alongside with the network connectivity. We also study the effects of setting a minimum rate for the key generation between nodes: only the edges that achieve a key rate larger than a positive threshold $ K_{\mathrm{min}} $ are considered a functional component of the network, and thus not pruned. This requirement is added to showcase more useful cases rather than a mere non-zero rate between nodes. However, even with this more demanding rule for pruning, the analysis conducted previously remains valid. 
While the connectivity curve unsurprisingly shifts, for densities above the percolation threshold the average key rate stays the same. This is an indication that most of the information exchanges happen through a small number of short, fast edges of the system, and that raising $ K_{\mathrm{min}} $ merely excludes some low-rate connections, unlikely to be part of the optimal paths chosen for communication. 

\begin{figure}[h!]
    \centering
    \begin{subfigure}[b]{0.5\textwidth}
        \centering
        \includegraphics[width=1\textwidth]{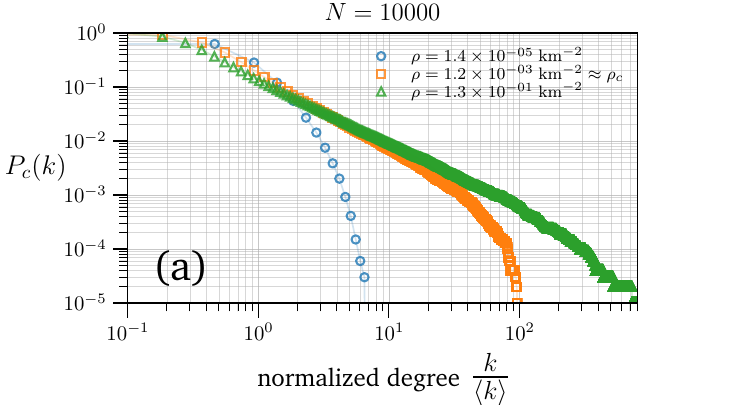}
        \phantomcaption
        \label{fig:deg_dist_hybrid}
    \end{subfigure}
    \vspace{-\myspace}
    \begin{subfigure}[b]{0.5\textwidth}
        \centering
        \includegraphics[width=\textwidth]{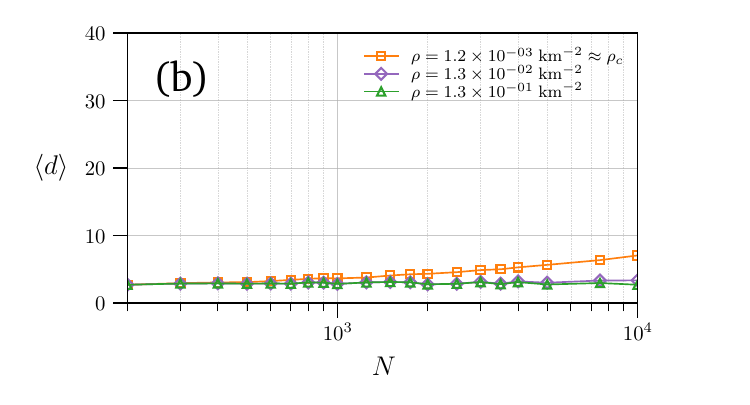}
        \phantomcaption
        \label{fig:distance_vs_logN_hybrid}
    \end{subfigure}
    \vspace{-\myspace}
    \begin{subfigure}[b]{0.5\textwidth}
        \centering
        \includegraphics[width=1\linewidth]{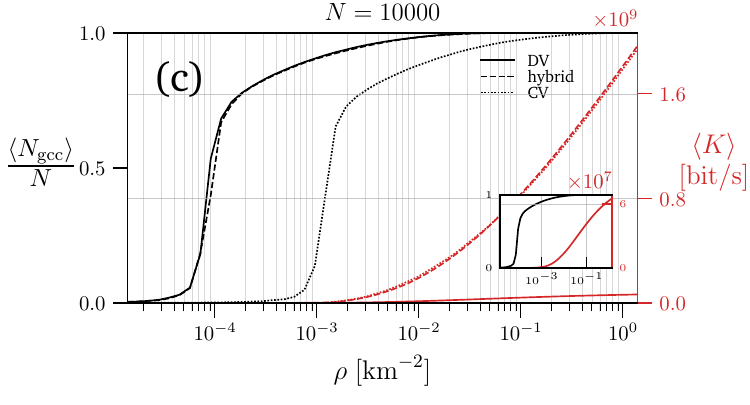}
        \phantomcaption
        \label{fig:DV_CV_comparison}
    \end{subfigure}
    \caption{
    (a) Complementary cumulative degree distribution $P_c(k)$ corresponding to the hybrid QKD protocol. When compared with~\ref{fig:deg_dist}, the hybrid networks exhibit a richer structure for each of the three densities, as appears from the tails of the distributions, that are non-zero for higher values of $\frac{k}{\langle k \rangle}$. (b) Average topological distance $\langle d \rangle$, as a function of the size $N$ of the network. The same densities and scale as in Fig.~\ref{fig:distance_vs_logN} are used to better highlight the improvement with respect to the CV-only scheme. (c) Performance of QKD networks based on our DV, CV and hybrid methods, in terms of both the relative size of the giant component $\frac{N_\mathrm{gcc}}{N}$ (black lines) and the average key rate $\langle K \rangle$ (red lines). To improve readability, the DV rate is replotted separately in the inset with an adjusted scale. The key rate $K$ that we associate to a generic pair of nodes $A$ and $B$ in the network is the one corresponding to the optimal route, among all the ones that connect $A$ and $B$. This route is found by minimizing the time $\Delta t$ taken for the generation of all the secret keys in each and every link along the route; it is then immediate to compute the key rate in bit/s as $K=\frac{1}{\Delta t}$. Considering that any QKD protocol can be independently and simultaneously executed on all links, $\Delta t$ is determined by the slowest link, as explained in detail in App.~\ref{App:figures_merit}. In order to compute $\langle K \rangle$, we repeated this procedure for many different $(A,B)$ pairs, over several realizations of the network.
    }
    \label{fig:hybrid_plots}
\end{figure}

The previous results show that, if  CV-QKD is used to prune links, some complex features are lost at low densities. They can be partially recovered through the integration of DV-QKD protocols in the model. We thus consider a hybrid model, where we select individually for each link the implementation that provides the highest rate between CV and DV. This implies that as density decreases and nodes get farther away from each other the share of DV links in the network grows.
The results are shown in Fig. ~\ref{fig:DV_CV_comparison}. The hybrid method allows us to get the best of both worlds: quantum-secure communication in long channels is restored and the percolation threshold is lowered by an order of magnitude, while a high average rate is maintained in densely populated systems where CV-QKD is used.
Even though the initial model, relying exclusively on CV protocols, was motivated by the high achievable rates and by the possibility of implementing it in the current classical Internet infrastructure, we note that this hybrid approach proves useful in recovering the properties of the initial Internet-like network (Fig.~\ref{fig:deg_dist_hybrid},~\ref{fig:distance_vs_logN_hybrid}).
Moreover, the security of a key generated between two distant nodes with our method requires that all nodes in the path between them are trusted. Reducing the number of such trusted intermediaries requires the recovery of small-worldness in the pruned network, and DV protocols help in that sense.

Our study can be expanded in many different directions. First of all, from a complex-network perspective, much larger-scale simulations will be required to confirm the value for the percolation threshold and the nature of the transition in Fig.~\ref{fig:connectivities_vs_rho}. This requires approaching the thermodynamic limit and therefore running simulations involving many more nodes. Another interesting avenue for further research is to employ real geolocation data, in order to build a network with a tighter correspondence to the current Internet infrastructure. From a QKD point of view, note that, in our analysis, the secret key between distant users was established through different trusted nodes. It would be interesting to adapt our framework to entanglement-based protocols, where nodes do not need to be trusted. Another possible extension is to incorporate satellite-based QKD links, which potentially enable any two users to perform secure quantum communications at very large distances without intermediaries. Finally, it is also relevant to consider the performance of other QKD protocols that require fewer assumptions for security, such as measurement-device-independent~\cite{mdiqkd} and device-independent~\cite{diqkd} QKD schemes. All these different relevant scenarios demonstrate that a lot remains to be done to fully understand the performance of QKD in real networked infrastructures and our work represents the first step in this direction.

\section*{Acknowledgements}
We thank Luis Trigo Vidarte for discussions about the experimental aspects of CV-QKD and Mari\'an Bogu\~n\'a for discussions on percolation in complex networks.

This work was supported by the European Union (ERC, AdG CERQUTE 834266; Horizon Europe, QSNP 101114043, QUCATS), the AXA Chair in Quantum Information Science, the Gobierno de España (Severo Ochoa CEX2019-000910-S, NextGenerationEU PRTR-C17.I1, FUNQIP and FPU predoctoral contract), Fundació Cellex, Fundació Mir-Puig, Generalitat de Catalunya (CERCA program) and the Italian Research Center on High Performance Computing, Big Data and Quantum Computing (through the European Union NextGenerationEU under Grant PUN:B93C22000620006), grant PID2022-137505NB-C22 funded by MCIN/AEI/10.13039/501100011033 and by ``ERDF A way of making Europe'', and the Generalitat de Catalunya grant number 2021SGR00856.

Views and opinions expressed are however those of the author(s) only and do not necessarily reflect those of the European Union or the European Research Council. Neither the European Union nor the granting authority can be held responsible for them.

\section*{Data and code availability}
The code developed for this study, along with the simulation results, is available in the GitHub repository \cite{githubQkdNetworkRepo}.

\noindent\makebox[\linewidth]{\rule{\columnwidth}{0.4pt}} 
\bibliographystyle{IEEEtran}
\bibliography{complex_nw_cvqkd}
\clearpage
\appendix
    \numberwithin{equation}{subsection}

\subsection{Network theory and geometric model}
\label{app:network}
\subsubsection{Definitions}
\label{app:networkdef}
Below, we go over some definitions of common concepts in network theory~\cite{Newman2010Networks}:
\begin{itemize}
    \item Depending on the symmetry or asymmetry of the connection between two nodes, the links within the graph are respectively said to be \textit{directed} or \textit{undirected}. Although many QKD protocols are asymmetric in their structure, they are primarily used in combination with the one-time pad cipher~\cite{pirandola2020advances} to attain unconditional security. Since this is a fundamentally symmetric cryptographic scheme, in this study of QKD networks we refer solely to \textit{undirected graphs}, where all edges are undirected.
    \item Two nodes sharing a link are said to be \textit{neighbors}; the number of neighbors of node $i$ takes the name of \textit{degree} of that specific node and is denoted as $k_i$.
    \item Graphs may be \textit{weighted} if every edge is associated with a number (or weight). In our model, each weight represents the secret-key rate in terms of bits per second.
    \item The \textit{connected components} (or simply \textit{components}) of a network form a partition of the network into disconnected groups of interconnected nodes. Consequently, a path exists between two nodes if and only if they belong to the same component.  In most real systems, the majority of nodes are contained in a single component, which is then called the \textit{giant component.} These systems are said to have \textit{percolated}. \\
\end{itemize}

\subsubsection{Real-world networks}
\label{subsubsec:real_networks}
Real-world networks have been studied with growing interest in the last decades in the context of complex network theory. Indeed, they manifest various complexity-related characteristics that have proven non-trivial to reproduce in theoretical models:
\begin{itemize}
    \item The \textit{small-world} property, which consists in the average shortest path length between two nodes growing slower than any polynomial of the network's size, i.e. the number $N$ of nodes \cite{ wattsCollectiveDynamicsSmallworld1998, chungAverageDistancesRandom2002, cohenScaleFreeNetworksAre2003, SmallWorldsBoguna2020}.
    \item The \textit{self-similarity} property, indicating invariance of the system under scale transformations \cite{ songSelfsimilarityComplexNetworks2005, garcia-perezMultiscaleUnfoldingReal2018}.
    \item The \textit{power-law} degree distribution: connected to self-similarity, this property implies that each node can have a number $k$ of neighbors that are distributed along an atypically broad interval. The degree distribution does not exhibit a well-defined variance and follows a power-law probability distribution \cite{ barabasiEmergenceScalingRandom1999,voitalovScalefreeNetworksWell2019}
    \begin{equation}\label{eq:powerlaw}
        P(k)\sim k^{-\gamma},
    \end{equation}
    usually with $2<\gamma<3$, which is the regime in which the variance of the distribution diverges. This implies that there exists a small number of nodes, called \textit{hubs}, that have a very large number of connections. Networks of this type are also called \emph{heterogeneous}, as opposed to \emph{homogeneous} networks with a narrow degree distribution.
    \item The presence of \textit{clustering} in the network, denoting the tendency of two neighbors of the same node to be themselves connected by a link, leading to triangle structures in the network.
\end{itemize}

\subsubsection{Geometric models}
\label{sec:geomodel}
\color{black}

In this article, we use the $\mathbb{S}^D$-model which takes into account both similarity and popularity~\cite{SelfSimilaritySerrano2008,SmallWorldsBoguna2020}. In this model, each node has a position $\{\vec{x}_i\}_{i=1}^N$ in a similarity space, which in this case is given by the $D$-dimensional hypersphere. The popularity of a node $i$ is encoded by a hidden degree $\kappa_i$, which can be shown to be equal to its corresponding expected degree. Each node $i$, thus, has a set of hidden parameters $\{\vec{x}_i,\kappa_i\}$. Each pair of nodes is connected with a probability 
\begin{equation}\label{eq:ConnectionProbability_0}
    p_{ij} = \frac{1}{1+\left(\frac{||\vec{x}_i-\vec{x}_j||}{(\mu\kappa_i\kappa_j)^{\frac{1}{D}}}\right)^\beta},
\end{equation}
where $\mu$ and $\beta$ tune the average degree and level of clustering, respectively (some typical values of $\mu$ and $\beta$ in embeddings of real-world networks can be found in \cite{Jankowski2023TheDM}). Note that this functional form is similar to a gravity law and leads to high connection probabilities when the inter-node distance $||\vec{x}_i-\vec{x}_j||$ is small or when the hidden degrees $\kappa_i,\kappa_j$ are large. 

It is important for our purposes that this model has an explicit geometric component. Key rates in QKD strongly depend on the physical distance between communicating nodes, which implies that topology alone is not sufficient to study the properties of a QKD network: it is required to know the coordinates of the agents in the embedding space.

To build a generative model for realistic complex networks, the $\mathbb{S}^2$ model is chosen. In this particular case, the similarity space is the surface of a sphere. As per the theoretical model, the radius of the sphere is assigned a value of $R_\mathrm{latent}=\sqrt{N/4\pi}$ to normalize the node density in the latent space to the unit. Regarding the positions of nodes in the real space, each node is assigned \say{geographical} coordinates that are equal, apart from a scaling factor, to the latent coordinates.
This choice is backed up by the fact that when embedding explicitly geometric real networks into the $\mathbb{S}^D$ model, nodes with similar hidden coordinates also tend to lie close together in real space~\cite{garcia-perezMercatorUncoveringFaithful2019}.
Finally, the value of the radius $R_\mathrm{real}$ of the real space represents a free parameter of our model, which we vary to control the density $\rho$ of the system.

\subsection{QKD rates}\label{App:qkd}

Here we describe the recipe used to assign secret-key rates to the links, or edges, in the considered QKD networks. We often use Alice and Bob to refer to the two nodes in a network link, as usually done in cryptography scenarios. First of all, quantum states are encoded on states of light that are produced with a repetition rate $\nu$, taken to be equal for all nodes. These light pulses propagate through channels corresponding to lossy fibers. For a channel, or link, connecting nodes $i$ and $j$, we employ the standard model for fiber losses in which the transmissivity of the channel, $T_{ij}$, is equal to
\begin{equation}\label{eq:Tij_exp}
    T_{ij} = 10^{- \alpha_\mathrm{att} L_{ij}/10}.
\end{equation}
Here, $\alpha_\mathrm{att}$ represents the attenuation coefficient at the channel per unit distance, and $L_{ij}$ is the physical distance between the two nodes.

We work in the asymptotic regime and then use the Devetak-Winter rate~\cite{devetak2005distillation} to bound the number of secret bits Alice and Bob generate per channel use. It reads
\begin{equation} \label{eq:DevetakWinter}
    K_\text{DW} = I(A:B) - \chi (B:E) ,
\end{equation}
that is, it estimates the secret key rate by comparing the mutual information $I(A:B)$ of Alice and Bob against Eve's information $E$, expressed in terms of the Holevo bound $\chi (B:E)$. Particularly, this expression refers to reverse reconciliation (i.e. Bob distills the final secret key, and sends error-correcting information to Alice), which is known to provide better results for CV-QKD \cite{navascuesSecurityBoundsContinuous2005,Silberhorn_2002} compared to direct reconciliation. For DV protocols, both direct and reverse reconciliation provide the same results. The exact expression for $K_\text{DW}$ depends on the considered protocol and is given next.

\subsubsection{Continuous-Variable QKD rates}\label{App:CVQKD}
We consider the standard Gaussian modulated CV-QKD protocol in which Alice prepares coherent states according to a Gaussian distribution centered at the phase-space origin and with modulation equal to $\sigma_A$. These states are sent to Bob who performs a homodyne measurement of one of the two light quadratures, $q$ or $p$. Alice and Bob can use the correlated information of the state prepared by Alice, denoted by $A$, and the measurement result by Bob, denoted by $B$, to establish the secret key, as proposed in~\cite{grosshans2002continuous}.

We consider a typical scenario where nodes are connected by additive white Gaussian noise channels characterized by a transmissivity $T$, see Eq.~\eqref{eq:Tij_exp}, and excess noise $\varepsilon$. 
We compute the value of the excess noise at the edge of the network connecting nodes $i$ and $j$ via the formula \cite{Luis_HighRateCVQKD,trigovidarte_2019}
\begin{equation}\label{eq:eps_ij}
	\varepsilon_{ij}=\varepsilon_0  \tau/(\eta T_{ij})\, 
\end{equation}
where $\varepsilon_0$ is the baseline excess noise (i.e., at Alice's side), whose effect is amplified by the detector efficiency $\eta$ and the transmissivity $T_{ij}$ of the channel. The term $\tau$ depends on the type of measurement, such that $\tau=1$ for homodyne and $\tau=2$ for heterodyne measurements. In this work, we study only homodyne measurements, i.e. we set $\tau=1$. The Devetak-Winter rate can be computed as a function of the two parameters $T$ and $\varepsilon$ as follows. 

For the given protocol, the covariance matrix $V_{AB}$ of the quadratures of $A$ and $B$ is equal to~\cite{navascuesSecurityBoundsContinuous2005}

\begin{equation} \label{eq:CovMatrix}
    \begin{pmatrix}
        \sigma^2_A \mathds{1} & \sqrt{T} (1+\sigma^2_A) \mathbf{Z} \\
        \sqrt{T} (1+\sigma^2_A) \mathbf{Z} & \left(T\sigma^2_A + 1 - T +\varepsilon T\right) \mathds{1}
    \end{pmatrix}.
\end{equation}

The mutual information is given by~\cite{pirandola_high-rate_2015}

\begin{equation}
    I(A:B) = \frac{1}{4}\log \left(\frac{{V}^q_{A}}{V^q_{A|q}}\right) + \frac{1}{4}\log \left(\frac{{V}^p_{A}}{V^p_{A|p}}\right).
\end{equation}
where $V^x_A=\sigma_A^2$ is the variance of quadrature $x\in \{q,p\}$ for Alice, and ${V}^x_{A|x}$ is the variance of quadrature $x$ for Alice conditioned on Bob's measurement. The latter can be obtained from the conditional covariance matrix ${V}_{A|x}$, computed as Schur's complement of $V_{AB}$ \cite{weedbrook_gaussian_2012}. If we write $V_{AB}$ in block form
\begin{equation}
    V_{AB} = \begin{pmatrix}
        V_A & C \\
        C^T & V_B
    \end{pmatrix},
\end{equation}
then Schur's complement is
\begin{equation} \label{eq:homodyne}
    {V}_{A|x} = V_{A} - C (\Pi_x V_B \Pi_x)^{-1} C^T.
\end{equation}
Here, $\Pi_q = \mathrm{diag}(1,0)$ and $\Pi_p = \mathrm{diag}(0,1)$. Note that $\Pi_x V_B \Pi_x$ will be singular, so $(\Pi_x V_B \Pi_x)^{-1}$ is a pseudoinverse.
On the other hand, the Holevo information can be expressed as \cite{Denys2021explicitasymptotic} 

\begin{equation}
    \chi (B:E) = g\left( \frac{\gamma_1-1}{2}\right) + g\left( \frac{\gamma_2-1}{2}\right) - g\left( \frac{\gamma'-1}{2}\right).
\end{equation}
where

\begin{equation}
    g(x):= (x+1)\log(x+1) - x\log(x)
\end{equation}
and $\{\gamma_1, \gamma_2\}$ are the symplectic eigenvalues \cite{weedbrook_gaussian_2012} of $V_{AB}$, whereas $\{\gamma'\}$ is the symplectic eigenvalue of ${V}_{A|x}$. \\

The final expression for the key rate associated to connection $(i,j)$ reads
\begin{equation}
\label{eq:cvrate}
    K_{ij}=\nu K_\text{DW} .
\end{equation}
In what follows we work with fixed values for the repetition rate, attenuation, detection efficiency and baseline excess noise and, therefore, the rate $K_{ij}$ only depends on the physical distance $L_{ij}$.\\

\subsubsection{Discrete Variable QKD rates}\label{Subsub:DVQKD}

To model DV-QKD links in the network, we consider the well-known BB84 protocol \cite{BB84}. To be more precise, we use a realization of BB84 based on single-photon states, where the information is encoded in the polarization of the photon. Similar performance is obtained for decoy-state protocols.\\

Single-photon states are sent via a fiber channel, again characterized by its transmissivity. For every round, Alice randomly chooses one out of four possible qubit states given by the $Z$ basis $\{\ket{0},\ket{1}\}$ or the $X$ basis $\{\ket{+},\ket{-}\}$, and sends it to Bob. On the same grounds, Bob randomly applies a measurement in one of the said bases for every round. From the measurement results, Alice and Bob form their classical key registers $A$ and $B$, respectively, which will be in disagreement when the bases are the same with a probability given by the \textit{Quantum Bit Error Rate} (QBER). In this case, the Devetak-Winter rate can be simplified~to~\cite{RajaThatCoolGuy_2019, Shor_SecProof_2000}

\begin{equation}
    K_\text{DW} = 1 -  h(Q_x) -  h(Q_z),
\end{equation}
where $h(.)$ represents the Shannon binary entropy, and  $Q_x$ and $Q_z$ are the QBER in either the $X$ or $Z$ basis. Although the two conjugated bases can, in general, have different QBERs, we can without loss of generality set the lowest to be equal to the highest and use
\begin{equation}
    K_\text{DW} = 1 -  2h(Q),
\end{equation}
where $Q=\max(Q_x,Q_z)$, since this is a valid lower bound on the secret-key rate. \\

In order to find a realistic value for the QBER, we consider a model~\cite{Shibata_2014} that takes into account the inefficiencies of the channel by considering \textit{dark counts}, which are the clicks on the detector that do not come from actual signals. This is done by splitting Bob's probability $P$ of observing a click event into
\begin{equation}
    P = P_\mathrm{s} + P_\mathrm{d}
\end{equation}
where $P_\mathrm{s}$ is the probability of a signal causing a click, whereas $P_\mathrm{d}$ is the probability of observing a click due to a dark count. With these terms, the QBER adjusted to the dark counts of the channel is

\begin{equation}
\label{eq:qber}
    Q = Q_0 \frac{P_\mathrm{s}}{P} +  \frac{P_\mathrm{d}}{2P}.
\end{equation}
Here, $Q_0$ represents the baseline QBER and we note that the second term of~\eqref{eq:qber} is multiplied by $1/2$ since dark counts provide a random outcome, and thus a bit in disagreement between Alice and Bob only half of the times. Let us elaborate on the probabilities by decomposing $P_\mathrm{s}$ as
\begin{equation}
    P_\mathrm{s} = \tilde{\nu} p_{\rm det} T,
\end{equation}
where $\tilde{\nu}$ and $p_{\rm det}$ are the efficiencies of the source and detection setups, respectively, and $T$ the transmissivity of the fiber channel. On the other hand, $P_\mathrm{d}$ scales as
\begin{equation}
    P_\mathrm{d} = R_\mathrm{d} \delta_\mathrm{d}
\end{equation}
with $R_\mathrm{d}$ the dark count rate, and $\delta_\mathrm{d}$ the detection window for Bob's detector.

The final expression for the key rate associated to connection $(i,j)$ reads
\begin{equation}
\label{eq:dvrate}
    K_{ij}=\nu P_\mathrm{s} K_\text{DW} .
\end{equation}
Again, this rate only depends on the physical distance $L_{ij}$ because the rest of parameters are fixed and equal for all the nodes. Apart from the fact that the expression for $K_\text{DW}$ varies, the difference with respect to the continuous-variable counterpart~\eqref{eq:cvrate} comes from the fact that single photons are detected with probability $P_S$, while a measurement outcome is always obtained in CV-QKD, in other words $P_\mathrm{s}=1$.\\

\subsection{Construction of an Internet-like QKD model}
\label{App:NetworkGen}
We embed an existing dataset of Autonomous Systems (dataset AS-733 taken from the Stanford Large Network Dataset Collection, representing an Autonomous System graph from January 2 2000~\cite{StanfordDataset}) in the $\mathbb{S}^2$ model through the software D-Mercator from \cite{garcia-perezMercatorUncoveringFaithful2019,Jankowski2023TheDM}.
From such set, we can extract the parameters $\beta$, $\mu$ of the $\mathbb{S}^2$ model, as explained in Sec.~\ref{sec:geomodel}. 
The obtained values for these quantities, namely $\beta=2.6261$ and $\mu=0.0233$, are used to generate networks with the connection probability given in eq.~\eqref{eq:ConnectionProbability_0}. This ensures having realistic levels of clustering and average degrees, respectively. Regarding the values of $\vec{x}_i$ and $\kappa_i$, although D-Mercator also returns a set of inferred coordinates for the embedded network, we choose to sample \say{synthetic} coordinates from appropriate probability distributions. This approach has two advantages: (a) it does not impose a constraint on the size of the network and (b) it allows for better randomization, avoiding projecting patterns of the training dataset onto the generative model. The rest of the procedure to build a quantum network is given as follows. 

\begin{scheme}
    \begin{enumerate}
	\item Create a set of $N$ uniformly distributed points on the unit sphere. From the coordinates $\vec{u}$ of these points, we derive both the latent coordinates $\vec{x}\equiv R_{\mathrm{latent}}\cdot\vec{u}$ and the real coordinates $\Vec{X}\equiv R_{\mathrm{real}}\cdot\vec{u}$.
    \item Sample a set of $\kappa$ coordinates from a power-law distribution $P(\kappa)\sim \kappa^{-\gamma}$: they will be the hidden degrees of the generative model. We set $\gamma=2.3$, compatibly with known values~\cite{faloutsosPowerlawRelationshipsInternet1999,barabasiEmergenceScalingRandom1999}.
	\item For every pair of nodes $(i,j)$, obtain the inter-node geodesic distances $||\vec{x}_i-\vec{x}_j||$: in the $\mathbb{S}^2$ model, it is the length of the shortest line connecting $i$ and $j$ on the surface of the latent sphere.
    \item Compute the connection probability for the nodes $(i,j)$ via the formula \eqref{eq:ConnectionProbability_0} for $p_{ij}$.
    \item For all the possible $O(N^2)$ couples of nodes, connect each pair $(i,j)$ with probability $p_{ij}$. 
    \item If any node is disconnected from the giant component, remove it and repeat the procedure until reaching the desired size $N$.
\end{enumerate}
\end{scheme}

After following these steps, we can study how employing a quantum communication setup between the nodes alters the structure of the network. For this, we model the edges as optical fibers that can be used to perform the Gaussian CV-QKD protocol described in Appendix~\ref{App:CVQKD}. 

In particular, we assign the secret-key rate $K_{ij}$~\eqref{eq:cvrate} to the edge connecting node $i$ and $j$, directly dependent on the distance $L_{ij}$, which is calculated with respect to the real coordinates $\vec{X}$. This also sets a critical distance over which no positive secret key rate is achievable. All the edges that exceed said distance are removed from the CV-QKD network. 

To improve the performance of the QKD network model, the \textit{hybrid} approach mentioned in Figure~\ref{fig:DV_CV_comparison} also employs DV-QKD connections. In this approach, each link implements either the CV- or the DV-QKD protocols, described in Appendix~\ref{App:qkd}, depending on which provides a higher key rate. Given the parameters listed in Table~\ref{tab:baselineparameters} of App.~\ref{app:AnalysisQNet}, and consistently with experimental results, DV-QKD is then preferred for long-range connections. This permits the reintegration, into the hybrid network, of many links that would be unusable in a purely CV-QKD system. 

In both cases, after the process of removing useless links, which is referred to as \textit{pruning} in the main text, the resulting graph has a connectivity that is below or equal to the one of the original graph. We are left with a network of active quantum channels for QKD each weighted with the corresponding secret key rate, thus exhibiting a different topology from the original \say{Internet} network.
We can then analyze the complex network properties and QKD performances of the resulting graph. This provides insight into the performance of a QKD network that would use the current Internet topology.

\subsection{Figures of merit}
\label{App:figures_merit}

In this work, most figures of merit are represented as functions of the node density in the real space $\rho=\frac{N}{4\pi R_{\mathrm{real}}^2}$. It is important to notice that, as the number of nodes increases, the computational effort becomes quite demanding. For this reason, in order to span a wider range of values for $\rho$, the parameter that we vary is $R_{\mathrm{real}}$, while keeping $N$ constant. However, for every computed quantity we also display the curves for a few different values of $N$, to control the impact of finite-size effects.

Let us now track the main properties derived from our QKD network architecture. The first point of interest is the \textit{connectivity}, defined as the ratio of the average size $\langle N_\mathrm{gcc} \rangle$ of the giant component with respect to the total number of nodes in the network. In particular, we consider a connectivity normalized to the unit. This means that fully disconnected graphs provide a value of $\frac{1}{N}$ while fully connected graphs, where a single giant component includes every node in the network, give a value 1. This is true in this work for systems with a very large $\rho$. 

\begin{figure}[h!]
    \centering
    \includegraphics[width=0.98\linewidth]{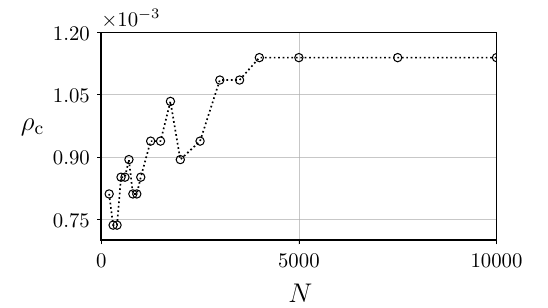}
    \caption{Critical density $\rho_c$ as a function of the size $N$ of the network. $\rho_c$, estimated as the position of the the peak of the susceptibility curve, appears to approach an asymptotic value.}
    \label{fig:rhoc_vs_N}
\end{figure}

The second property that we study is the ratio between variance and average of the size $N_\mathrm{gcc}$ of the giant component of the resulting network, which defines the \textit{susceptibility} $\chi$ \cite{ferreiraEpidemicThresholdsSusceptibleinfectedsusceptible2012, castellanoNumericalStudyPercolation2016}:
\begin{equation}
    \chi\equiv \frac{\langle (N_\mathrm{gcc} - \langle N_\mathrm{gcc} \rangle)^2 \rangle}{\langle N_\mathrm{gcc} \rangle}.
\end{equation}
As it is connected to fluctuations in the system, it is a good indicator of a percolation phase transition: the node density at which $\chi$ reaches a maximum identifies a critical density $\rho_c$ after which the network has a giant connected component after pruning. Analyzing how $\rho_c$ varies with $N$ (Fig. \ref{fig:rhoc_vs_N}) allows us to partially discriminate the contribution of finite-size effects.

Another quantity that we consider to gain insight into the impact of node density on the structure of the system is its degree distribution. Specifically, we show the \textit{complementary cumulative degree distribution} of the nodes $P_c(k)$ according to their corresponding degrees $k$. The term $P_c(k)$ is conventionally defined as the fraction of nodes in the network having $k$ or more connections: if $N_k$ is the number of nodes with degree $k$ then 

\begin{equation}
    P_c(k)\equiv 1-\frac{1}{N}\sum_{i=0}^{k-1} N_i.
\end{equation}

Finally, we monitor the average shortest path length, or topological distance, between nodes $\langle d\rangle$. In small-world networks, $\langle d\rangle$ is very small, even when the total amount of nodes is large. Typically, we call a network small-world when $\langle d\rangle$ scales with $N$ slower than any power-law. A typical example is given by the Erdos-Renyi graph where $\langle d\rangle\sim \log N$. 
The $\mathbb{S}^2$ model can be shown to produce \emph{ultra-small} worlds when the degree distribution is taken to be heterogeneous, i.e., when $2<\gamma<3$~\cite{SmallWorldsBoguna2020,cohenScaleFreeNetworksAre2003}. Here the average shortest paths are extremely small, namely, $\langle d\rangle\sim \log \log N$. This is due to the presence of hubs, which are connected to a large amount of the nodes in the network, therefore leading to many shortcuts that reduce $\langle d\rangle$. \\

From the perspective of quantum communications, we consider that the creation of a secret key between any two nodes $A$ and $B$ is done through a series of point-to-point secret key exchanges, along the most effective path $\mathcal{P}^*$ linking these two nodes. The parameter of interest that we define is thus the \textit{secret key rate} $ K $ achievable along $\mathcal{P}^*$, which is found through a path-optimization algorithm. In general, such an algorithm involves the minimization (or maximization, as in our case) of a certain cost function $W(\mathcal{P})$ over the set $\{\mathcal{P}\}$ of all possible paths connecting $A$ and $B$. 
\begin{equation}
    \mathcal{P}^* \equiv \mathrm{arg}\max_{\{\mathcal{P}\}} W(\mathcal{P})
\end{equation}
Here $W(\mathcal{P})$ is the secret key rate associated with the path $ \mathcal{P}$. In this work, we consider that the key distributions across the links included in $\mathcal{P}$ all happen in parallel. Hence, $W(\mathcal{P})$ is the minimum key rate among the key rates of the direct connections that $ \mathcal{P}$ consists of:
\begin{equation}
    W(\mathcal{P}) = \min_{(i,j)\in \mathcal{P}} K_{ij}\ ,
\end{equation}
where given the length $L_{ij}$ of the connection $(i,j)$, the corresponding rate $K_{ij}$ is given by Eqs.~(\ref{eq:cvrate},\ref{eq:dvrate}). The slowest channel (the one with the lowest key rate) then represents the bottleneck for the protocol, thus setting an upper bound on the achievable rate across the whole connection. Once we obtain $\mathcal{P}^*$, we take the corresponding weight as the key rate for the pair $(A,B)$: 
\begin{equation}
    K_{AB} = W(\mathcal{P}^*) \equiv \max_{\{\mathcal{P}\}} \min_{(i,j)\in \mathcal{P}} K_{ij}
\end{equation}
The shortest-path algorithm used, which may be viewed as a variation of Dijkstra's algorithm \cite{DijkstraAlg1959}, always finds the optimal solution when available. It returns the path and the corresponding key rate, in bits per second. When no path is available between the nodes, the rate is set to zero. In order to gain statistical significance, the procedure is repeated, and the results averaged, over a large number of pairs $(A,B)$ for any instance of the quantum network, to obtain an average key rate $\langle K \rangle$.

\subsection{Numerical implementation} \label{app:AnalysisQNet}

Given the procedure described in the previous Sections, a numerical analysis of the model can be carried by using the Python package NetworkX \cite{NetworkX_2008} and the software Mercator \cite{Jankowski2023TheDM}. The code is available on the Github repository \cite{githubQkdNetworkRepo}, and the parameters used in the simulation are given in Table~\ref{tab:baselineparameters}.\\

\begin{table}[!ht]
    \centering
\begin{tabular}{l|c|l}
    Symbol & Value & Description  \\ \hline
    $\alpha_\mathrm{att}$ & $0.18$ dB/km & Fiber loss per kilometer\\
    $\nu$ & $1$ GHz & Repetition rate \\
    $\varepsilon_0$ & $0.005$ SNU & Excess noise (CV-QKD) \\ 
    $\eta$ & $0.8$  & Detector efficiency (CV-QKD)  \\
    $\sigma_A$ & $10^2$ & Alice's modulation (CV-QKD)\\
    $p_{\rm det}$ & $0.95$  & Detector efficiency (DV-QKD) \\
    $\tilde{\nu}$ & $ 0.1$ & Source efficiency (DV-QKD) \\
    $R_\mathrm{d}$ & $100$ Hz & Dark count rate (DV-QKD) \\
    $\delta_\mathrm{d}$ & $100$ ps& Detection window (DV-QKD) \\
    $Q_0$ & 1\% & Baseline QBER (DV-QKD)
\end{tabular}
    \caption{Baseline simulation parameters. When relevant, it is indicated in parenthesis if the parameter corresponds to the CV- or DV-QKD protocol.}
    \label{tab:baselineparameters}
\end{table}
As the positions and node degrees are randomly distributed every time a network is generated due to the non-deterministic nature of our generative model, we average (unless otherwise stated) over 10 instances for a fixed value of $\rho$. This allows us to gather enough statistical evidence to extract meaningful results.

Although complex behaviors can already be observed in relatively small networks ($N\approx 100$), it is more convenient to simulate larger systems to neglect finite-size effects. However, both the routine for building the quantum network and the algorithm for the optimal path scale as $O(N^2)$. For a practical study that balances the numerical performance of the code and the reliability of the results, we set $N \in [200,10000]$. Note that the number of functional edges, as well as the fraction of completely disconnected nodes after pruning, depends dramatically on the density of the nodes.

\noindent\makebox[\linewidth]{\rule{\columnwidth}{0.4pt}}

\end{document}